\documentstyle[twoside,wrapfig,epsf]{article}

\catcode`\@=11
\long\def\@makefntext#1{
\protect\noindent \hbox to 3.2pt {\hskip-.9pt
$^{{\eightrm\@thefnmark}}$\hfil}#1\hfill}               

\def\thefootnote{\fnsymbol{footnote}}
\def\@makefnmark{\hbox to 0pt{$^{\@thefnmark}$\hss}}    

\def\ps@myheadings{\let\@mkboth\@gobbletwo
\def\@oddhead{\hbox{}
\rightmark\hfil\eightrm\thepage}
\def\@oddfoot{}\def\@evenhead{\eightrm\thepage\hfil
\leftmark\hbox{}}\def\@evenfoot{}
\def\sectionmark##1{}\def\subsectionmark##1{}}

\oddsidemargin=\evensidemargin
\renewcommand{\thefootnote}{\fnsymbol{footnote}}

\newcounter{sectionc}\newcounter{subsectionc}\newcounter{subsubsectionc}
\renewcommand{\section}[1] {\vspace{12pt}\addtocounter{sectionc}{1}
\setcounter{subsectionc}{0}\setcounter{subsubsectionc}{0}\noindent
        {\tenbf\thesectionc. #1}\par\vspace{5pt}}
\renewcommand{\subsection}[1] {\vspace{12pt}\addtocounter{subsectionc}{1}
        \setcounter{subsubsectionc}{0}\noindent
        {\bf\thesectionc.\thesubsectionc. {\kern1pt \bfit #1}}\par\vspace{5pt}}
\renewcommand{\subsubsection}[1] {\vspace{12pt}\addtocounter{subsubsectionc}{1}
        \noindent{\tenrm\thesectionc.\thesubsectionc.\thesubsubsectionc.
        {\kern1pt \tenit #1}}\par\vspace{5pt}}
\newcommand{\nonumsection}[1] {\vspace{12pt}\noindent{\tenbf #1}
        \par\vspace{5pt}}

\newcounter{appendixc}
\newcounter{subappendixc}[appendixc]
\newcounter{subsubappendixc}[subappendixc]
\renewcommand{\thesubappendixc}{\Alph{appendixc}.\arabic{subappendixc}}
\renewcommand{\thesubsubappendixc}
        {\Alph{appendixc}.\arabic{subappendixc}.\arabic{subsubappendixc}}

\renewcommand{\appendix}[1] {\vspace{12pt}
        \refstepcounter{appendixc}
        \setcounter{figure}{0}
        \setcounter{table}{0}
        \setcounter{lemma}{0}
        \setcounter{theorem}{0}
        \setcounter{corollary}{0}
        \setcounter{definition}{0}
        \setcounter{equation}{0}
        \renewcommand{\thefigure}{\Alph{appendixc}.\arabic{figure}}
        \renewcommand{\thetable}{\Alph{appendixc}.\arabic{table}}
        \renewcommand{\theappendixc}{\Alph{appendixc}}
        \renewcommand{\thelemma}{\Alph{appendixc}.\arabic{lemma}}
        \renewcommand{\thetheorem}{\Alph{appendixc}.\arabic{theorem}}
        \renewcommand{\thedefinition}{\Alph{appendixc}.\arabic{definition}}
        \renewcommand{\thecorollary}{\Alph{appendixc}.\arabic{corollary}}
        \renewcommand{\theequation}{\Alph{appendixc}.\arabic{equation}}
        \noindent{\tenbf Appendix \theappendixc #1}\par\vspace{5pt}}
\newcommand{\subappendix}[1] {\vspace{12pt}
        \refstepcounter{subappendixc}
        \noindent{\bf Appendix \thesubappendixc. {\kern1pt \bfit #1}}
        \par\vspace{5pt}}
\newcommand{\subsubappendix}[1] {\vspace{12pt}
        \refstepcounter{subsubappendixc}
        \noindent{\rm Appendix \thesubsubappendixc. {\kern1pt \tenit #1}}
        \par\vspace{5pt}}

\newcommand{\textlineskip}{\baselineskip=13pt}
\newcommand{\smalllineskip}{\baselineskip=10pt}

\def\eightcirc{
\begin{picture}(0,0)
\put(4.4,1.8){\circle{6.5}}
\end{picture}}
\def\eightcopyright{\eightcirc\kern2.7pt\hbox{\eightrm c}}

\newcommand{\copyrightheading}[1]
        {\vspace*{-2.5cm}\smalllineskip{\flushleft
        {\footnotesize $\eightcopyright$\, World Scientific Publishing
         Company}\\
         }}

\def\abstracts#1#2#3{{
        \centering{\begin{minipage}{4.5in}\baselineskip=10pt\footnotesize
        \parindent=0pt #1\par
        \parindent=15pt #2\par
        \parindent=15pt #3
        \end{minipage}}\par}}

\newcommand{\bibit}{\nineit}
\newcommand{\bibbf}{\ninebf}
\renewenvironment{thebibliography}[1]
        {\frenchspacing
         \ninerm\baselineskip=11pt
         \begin{list}{\arabic{enumi}.}
        {\usecounter{enumi}\setlength{\parsep}{0pt}
         \setlength{\leftmargin 12.7pt}{\rightmargin 0pt} 
         \setlength{\itemsep}{0pt} \settowidth
        {\labelwidth}{#1.}\sloppy}}{\end{list}}

\newcounter{itemlistc}
\newcounter{romanlistc}
\newcounter{alphlistc}
\newcounter{arabiclistc}

\newcommand{\fcaption}[1]{
        \refstepcounter{figure}
        \setbox\@tempboxa = \hbox{\footnotesize Fig.~\thefigure. #1}
        \ifdim \wd\@tempboxa > 5in
           {\begin{center}
        \parbox{5in}{\footnotesize\smalllineskip Fig.~\thefigure. #1}
            \end{center}}
        \else
             {\begin{center}
             {\footnotesize Fig.~\thefigure. #1}
              \end{center}}
        \fi}

\newcommand{\tcaption}[1]{
        \refstepcounter{table}
        \setbox\@tempboxa = \hbox{\footnotesize Table~\thetable. #1}
        \ifdim \wd\@tempboxa > 5in
           {\begin{center}
        \parbox{5in}{\footnotesize\smalllineskip Table~\thetable. #1}
            \end{center}}
        \else
             {\begin{center}
             {\footnotesize Table~\thetable. #1}
              \end{center}}
        \fi}

\def\@citex[#1]#2{\if@filesw\immediate\write\@auxout
        {\string\citation{#2}}\fi
\def\@citea{}\@cite{\@for\@citeb:=#2\do
        {\@citea\def\@citea{,}\@ifundefined
        {b@\@citeb}{{\bf ?}\@warning
        {Citation `\@citeb' on page \thepage \space undefined}}
        {\csname b@\@citeb\endcsname}}}{#1}}

\newif\if@cghi
\def\cite{\@cghitrue\@ifnextchar [{\@tempswatrue
        \@citex}{\@tempswafalse\@citex[]}}
\def\citelow{\@cghifalse\@ifnextchar [{\@tempswatrue
        \@citex}{\@tempswafalse\@citex[]}}
\def\@cite#1#2{{$\null^{#1}$\if@tempswa\typeout
        {IJCGA warning: optional citation argument
        ignored: `#2'} \fi}}

\def\pmb#1{\setbox0=\hbox{#1}
        \kern-.025em\copy0\kern-\wd0
        \kern.05em\copy0\kern-\wd0
        \kern-.025em\raise.0433em\box0}

\def\fnt#1#2{\footnotetext{\kern-.3em
        {$^{\mbox{\scriptsize #1}}$}{#2}}}

\def\fpage#1{\begingroup
\voffset=.3in
\thispagestyle{empty}\begin{table}[b]\centerline{\footnotesize #1}
        \end{table}\endgroup}

\def\runninghead#1#2{\pagestyle{myheadings}
\markboth{{\protect\footnotesize\it{\quad #1}}\hfill}
{\hfill{\protect\footnotesize\it{#2\quad}}}}
\font\tenrm=cmr10
\font\tenit=cmti10
\font\tenbf=cmbx10
\font\bfit=cmbxti10 at 10pt
\font\ninerm=cmr9
\font\nineit=cmti9
\font\ninebf=cmbx9
\font\eightrm=cmr8

\def\qed{\hbox{${\vcenter{\vbox{                        
   \hrule height 0.4pt\hbox{\vrule width 0.4pt height 6pt
   \kern5pt\vrule width 0.4pt}\hrule height 0.4pt}}}$}}

\renewcommand{\thefootnote}{\fnsymbol{footnote}}        

\def\bsc{{\sc a\kern-6.4pt\sc a\kern-6.4pt\sc a}}       
\def\bflatex{\bf L\kern-.30em\raise.3ex\hbox{\bsc}\kern-.14em
T\kern-.1667em\lower.7ex\hbox{E}\kern-.125em X}

\def\vsp{\noalign{\vskip 0.3cm}}
\def\be{\begin{equation}}
\def\ee{\end{equation}}
\def\bea{\begin{eqnarray}}
\def\eea{\end{eqnarray}}
\def\hlf{{1\over 2}}
\def\ve{\varepsilon}
\begin{document}

\runninghead{Instructions for Typesetting Camera-Ready
Manuscripts} {Instructions for Typesetting Camera-Ready
Manuscripts}

\normalsize\textlineskip
\thispagestyle{empty}
\setcounter{page}{1}

\copyrightheading{}                     

\vspace*{0.88truein}
\fpage{1}
\centerline{\bf Automatic Calculation of 2-loop Weak Corrections}
\vspace*{0.035truein}
\centerline{\bf to Muon Anomalous Magnetic Moment \footnote{
Talk presented in AIHENP95 at Pisa, April 1995.}}
\vspace*{0.32truein}
\centerline{\footnotesize Toshiaki KANEKO \footnote{
Present Address: LAPP, IN$^2$P$^3$,Annecy,CNRS, France}}
\vspace*{0.017truein}
\centerline{\footnotesize\it Faculty of General Education,
 Meijigakuin University }
\baselineskip=10pt
\centerline{\footnotesize\it Totsuka,Yokohama 244 , JAPAN }
\vspace*{10pt}
\centerline{\normalsize and}
\vspace*{10pt}
\centerline{\footnotesize Nobuya NAKAZAWA}
\vspace*{0.017truein}
\centerline{\footnotesize\it Department of Physics, Kogakuin University}
\baselineskip=10pt
\centerline{\footnotesize\it
Nishi-Shinjuku 1-24, Shinjuku, Tokyo 160, Japan}
\vspace*{0.125truein}
\centerline{\footnotesize Received ( April 6, 1995 ) }
\vspace*{0.11truein}
\abstracts{
An automatic system to calculate
two loop weak corrections to muon anomalous magnetic moment
is discussed. Diagrams are classified into eight types,
according to their topology.
Adopting Civitanovi\'c-Kinoshita
representation of Feynman amplitude and using the
topological property, the renormalization is
performed consistently by $n$-dimensional regularization method.
}{}{}



\vspace*{1pt}\textlineskip      
\section{Introduction}          
\vspace*{-0.5pt}
\noindent
GRACE system$^1$ has been extended to produce one loop Feynman
amplitude in automatic way and the order $\alpha$ corrections
to several $2 \rightarrow 2 $ reactions are calculated$^2$.
The GRACE can also generate higher loop diagrams and correponding
amplitudes. As a first application of the system to two-loop
calculation, we consider the weak corrections to muon anomalous
magnetic moment.\\
 QED corrections to muon ($g$-2) are calculated up to eighth order
and its valus is$^3$
$$
\Delta a_{QED}^{(8)}=(g-2)/2 = 1165846947(\pm 46 \pm 28)\times
10^{-12}
$$
The first error arises from theoretical uncertainty and
the second one from a measurement uncertainty of
the fine structure constant $\alpha$.
The next order corrections in QED are estimated as
$\Delta a_{QED}^{(10)} = (39\pm 10)\times 10^{-11}$.
The vacuum polarization due to hadrons contributes
$
\Delta a_{hadron}=(703\pm 19)\times 10^{-10}.
$
The weak interaction calculated at one loop level
contribute
$
\Delta a_{Weak}^{1-loop} = (195\pm 1)\times 10^{-11}.
$
The two loop weak correction is estimated in the
leading logarithmic approximation and it reaches
about 22 \% of the one loop result$^4$.
So, it is worth to get two-loop contributions
from 1678 diagrams completely in automatic way.

\section{Formalism}
We adopt the Civitanovi\'c-Kinoshita$^5$
 representation of the Feynman
amplitude. \\
Combining the denominators by Feynman parameters,
the general muon vertex at two loop level
(in the case of 6 internal lines)
 is expressed as
\be
\Gamma_\mu
= \Gamma(6)\int
 \prod dz_j \delta(1-\sum_j z_j)
 \int
 \frac{d^n \ell_1}{(2\pi)^n}
 \frac{d^n \ell_2}{(2\pi)^n}
 \frac{F(D)}{\sum_j z_j(p_j^2-m_j^2)}
\ee
where, $p_j$  is a momentum flowing on the internal line $j$ and
expressed as
\be
p_j = \sum_{s=1,2} \eta_s(j)\ell_s + q_j.
\ee
The symbol $\eta_s(j)$ represents (1,-1,0) according to the flow of
loop momentum $\ell_s$ and $q_j$ is a external momentum
flow on the line $j$.
 The $F(D)$ represents a numerator and is expressed by several momenta
and $\gamma$ matrices.
By shifting the loop momenta to delete linear terms, the denominator
becomes
\be
{\rm Denom.} = \sum_{s,t}U_{st}(\ell_s\cdot\ell_t)-V
\ee
\be
V =\sum_j z_j m_j^2
-\sum_j z_j(q_j\cdot q_j)
+\frac{1}{{\rm detU}}\sum_{i,j}z_iz_jB_{ij}(q_i\cdot q_j),
\ee
where
\be
U_{s,t} = \sum_{j=1}^6 z_j\eta_s(j)\eta_t(j)
\ee
and
\be
B_{ij}=\sum_{s,t}\eta_s(i)\eta_t(j)U^{-1}_{st}\cdot{\rm detU} = B_{ji}.
\ee
These matrices depend only on the topology of the diagram.
By changing the basis of the independent loop momenta, the matrix $U$
is diagonalized and we rescale the loop momenta.
We adopt $n$-dimensional regularization method to remove the
Ultra Violet divergence. As the numerator can be reproduced
by the differential-integral operator $D_j^\mu$, we first
integrate over $n$-dimensional loop momenta.
The operator $D_j^\mu$ ($j$ represents the line number)
is defined as follows.
\be
D_j^\mu = \frac{1}{2}\int\nolimits_{m_j^2}^{\infty}
dm_j^2 \frac{\partial}{\partial q_{j\mu}}
\ee
The basic relation to pick up the numerator is
\be
D_j^\mu \frac{1}{(p_j^2 - m_j^2)}
=\frac{p_j^\mu}{(p_j^2 - m_j^2)}.
\ee
The fundamental relations necessary for us are given by
\be
D_j^\mu\frac{1}{V^m} =\frac{{Q'_j}^\mu}{V^m}
\ee
\be
D_i^\mu D_j^\nu \frac{1}{V^m}
=\frac{{Q'_i}^\mu{Q'_j}^\nu}{V^m}
+\left(- \frac{1}{2\;\rm detU}\right)\frac{g^{\mu\nu}}{(m-1)}
       \frac{B'_{ij}}{V^{m-1}}
\ee
where
\be
{Q'_j}^\mu = q_j^\mu -\frac{1}{\rm detU} \sum_i z_iB_{ij}q_i^\mu
= -\frac{1}{\rm detU}\sum_i z_i B'_{ij}q_i^\mu
\ee
The numerator can be reproduced by replacing $p_j^\mu$ with the operator
$D_j^\mu$. The basic formula for our calculation becomes
(except for coupling constant),

\bea
\Gamma^\mu
&=&\frac{1}{(4\pi)^n}\int
\prod dz_j \delta(1-\sum_j z_j)
\left[\frac{\Gamma(6-n)}{({\rm detU})^{n/2}}
\frac{F^\mu_0}{(V-i\varepsilon)^{6-n}}  \right.
\nonumber \\
\vsp
&+&\left. \frac{\Gamma(5-n)}{-2\;({\rm detU})^{n/2+1}}\frac{F^\mu_1}
{(V-i\varepsilon)^{5-n}}
+\frac{\Gamma(4-n)}{4\;({\rm detU})^{n/2+2}}\frac{F^\mu_2}
{(V-i\varepsilon)^{4-n}}
\right]
\eea

Practical method to generate proper numerator is as follow.
\begin{itemize}
\item
replace $p_j^\mu $ in numerator $\rightarrow (ck_j^\mu  +{Q'_j}^\mu) $
\item
$c^0$  terms $\rightarrow  F_0^\mu$
\item
$c^2$  terms $\rightarrow F_1^\mu$ :~~~~
replace $k_i^\mu k_j^\nu
\rightarrow (-1/(2\;{\rm detU}))B'_{ij}g^{\mu\nu} $
\item
$c^4$  terms $\rightarrow F_2^\mu$ : ~~~~replace \\
 $k_i^\mu k_j^\nu k_k^\lambda k_\ell^\sigma
 \rightarrow (-1/(2\;{\rm detU}))^2
\left\{
g_{\mu\nu}g_{\lambda\sigma}B'_{ij}B'_{k\ell}
+g_{\mu\lambda}g_{\nu\sigma}B'_{ik}B'_{j\ell}
+g_{\mu\sigma}g_{\nu\lambda}B'_{i\ell}B'_{jk} \right\}
$
\end{itemize}

In the above replacement, the contraction of Lorentz indices should be
done in $n$-dimension.
We introduce the following projection operator in 4-dimension
to pick up the contribution
to $(g-2)$.($q$:photon mom., $(p-q/2),(p+q/2)$:muon mom.)
\be
{\rm Proj}(\mu) = \frac{1}{4}(\rlap{/}p-\hlf\rlap{/}q+{\rm m})
    ({\rm m}\;\gamma_\mu (p.p) -({\rm m}^2+\frac{q.q}{2})p_\mu )
    (\rlap{/}p+\hlf\rlap{/}q + {\rm m})
\ee
By this projection, the $F_2$ term drops out.
We replace $F_0,F_1$ with the projected one $f0,f1$. The factor
$(-1/2)$ is included in $f1$.

\section{Removal of the Ultra Violet Divergence}

The ultra violet$\;$(UV) divergence arises from the one loop sub
diagram$^6$.
The over all divergence ($F_2$ term) is irrerevant in our case,
because of the projection. It is necessary for order $(\alpha^2)$
charge  renormalization.
The diagrams are classified into 8 different types of topology.
They are shown in the Fig.1.
 We will discuss two types of topology as examples. The vertex and
 self-energy type.(See Fig.2.)

\subsection{Vertex type}
\begin{figure}
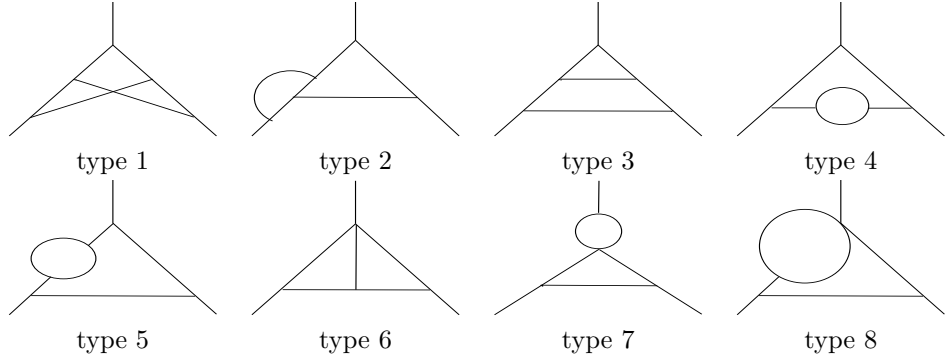

   \begin{center}
   \begin{tabular}{cccc}
       {\def\epsfsize#1#2{0.27#1\epsfysize0.2#2}\epsfbox{gr1.eps}} &
       {\def\epsfsize#1#2{0.27#1\epsfysize0.2#2}\epsfbox{gr2.eps}} &
       {\def\epsfsize#1#2{0.27#1\epsfysize0.2#2}\epsfbox{gr3.eps}} &
       {\def\epsfsize#1#2{0.27#1\epsfysize0.2#2}\epsfbox{gr4.eps}} \\
type 1 & type 2 & type 3 & type 4 \\
       {\def\epsfsize#1#2{0.27#1\epsfysize0.2#2}\epsfbox{gr5.eps}} &
       {\def\epsfsize#1#2{0.27#1\epsfysize0.2#2}\epsfbox{gr6.eps}} &
       {\def\epsfsize#1#2{0.27#1\epsfysize0.2#2}\epsfbox{gr7.eps}} &
       {\def\epsfsize#1#2{0.27#1\epsfysize0.2#2}\epsfbox{gr8.eps}} \\
type 5 & type 6 & type 7 & type 8
    \end{tabular} \end{center}
\caption{Types of Topology}
\end{figure}
We divide the diagram into a sub diagram ($S$) and
 the remained $(R)$.
The correponding Feynman parameters are
\be
(z_2,z_3,z_6) ~~~\in ~~S, ~~~~~~~~~
(z_1,z_4,z_5) ~~~\in ~~R
\ee
The determinant (detU) has a semi-factorized form in the
parameters.
This implies that the singularity occurs when all the
parameters in sub-diagram $S$ tends to 0.
In order to see this clearly,
we reparametrize the Feynman parameters as follows.
\be
z_2 = x(1-y),~~~~~ z_3 = x y (1-z), ~~~~~ z_6 = xyz
\ee
\be
z_1 = (1-x)(1-u),~~~~~ z_4 = (1-x)u(1-v), ~~~~~z_5 = (1-x)uv
\ee
The measure of integration is changed into
$x^2(1-x)^2yu$.
When $x$ approaches 0 then the detU also tends to 0.
This is the origin of divergence.
It is important that the structure of determinant only depends
on the topology of the diagram and the type of different topology
is restricted in small number. So we can prepare the
subtraction procedures beforehand,
according to the type of topology.
The UV divergence occurs from the second term.
\begin{itemize}
\item the second term $f1$
\end{itemize}
Setting $\varepsilon = 2 - n/2$,
the main part of the integrand becomes
\be
\Gamma(1+2\varepsilon)
\frac{x^2(1-x)^2yu}
{x^{3-\varepsilon}u(x,y,z)^3}
\frac{f10+\ve\cdot f11}
{V^{1+2\varepsilon}}
(4\pi)^{2\varepsilon} u(x,y,z)^\varepsilon
\ee
where ${\rm detU}=x\cdot u(x,y,z)$.
(We must keep numerators up to $O(\ve)$.)
We can see the existence of the singularity at
$\varepsilon = 0 $ when $x \rightarrow 0$,
due to the factor $x^{\varepsilon-1}$ .
In order to extract the singularity,
let us consider the following integral.
\be
I = \int\nolimits_0^1 x^{\varepsilon-1}
(h_0(x)+\ve\cdot h_1(x)) dx
\ee
Partial integfration and the expansion in $\ve$ give
\be
I = \frac{1}{\ve} h_0(0)+h_1(0)
-\int\nolimits_0^1 \log x \frac{d h_0(x)}{dx}
\ee
Using the above formula,
the contribution from $f1$ term
turns out (as the integrand of $dydzdudv\{yu\}$)
\bea
F^{(4)}_2(0) &=&
\left( \frac{1}{\ve} + 2\log (4\pi) -2 \gamma_E \right)
\frac{f10(0)}{V(0)}
+\frac{f10(0)}{V(0)}(-2\log V(0) )
+\frac{f11(0)}{V(0)} \nonumber \\
\vsp
&-& \int\nolimits_0^1 \log(x)
\frac{\partial}{\partial x}\left(
\frac{(1-x)^2}{u(x,y,z)^3}
\frac{f10(x)}{V(x)}\right)
\eea
  in unit of $(\alpha/\pi)^2/16$.

\subsection{Self energy type}
Similarly, we divide the diagram into a sub diagram ($S$) and
 the remained $(R)$.
The correponding Feynman parameters are
\be
(z_2,z_5) ~~~\in ~~S, ~~~~~~~~~
(z_1,z_3,z_4,z_6) ~~~\in ~~R
\ee
Reparametrization of Feynman parameters is done in the similar way.
\be
z_2 = x(1-y),~~~~~ z_5 = x y
\ee
\be
z_1 = (1-x)(1-u),~~~ z_3 = (1-x)u(1-v),
{}~~~z_4 = (1-x)uv(1-w),~~~~z_5=(1-x)uvw
\ee
The measure is slightly different from the previous case,
$x(1-x)^3u^2v$.
The singularity occurs even in the first term $f0$. The treatment
is the same as the $f1$ in the previous case.
The second term has the differnt structure.
The main part of the integrand is
\be
\Gamma(1+2\varepsilon)
\frac{x(1-x)^3u^2v}
{x^{3-\varepsilon}u(x,y)^3}
\frac{f10(x)+\ve\cdot f11(x)}
{V^{1+2\varepsilon}}
(4\pi)^{2\varepsilon} u(x,y)^\varepsilon .
\ee
To discuss the singularity, let
us consider the following integral.
\be
I = \int\nolimits_0^1 x^{\varepsilon-2}(g_0(x)+\ve\cdot g_1(x)) dx
\ee
Partial integration and the expansion in $\ve$ give
\be
I=
\frac{1}{\ve}\frac{dg_0(0)}{dx}
+\frac{dg_0(0)}{dx}+\frac{dg_1(0)}{dx}
-g_0(1)
 -\int\nolimits_0^1 \log x
 \frac{d^2 g_0(x)}{dx^2 }dx
\ee
By this formula, we can extract UV divergent part and the finite contribution.
\begin{figure}
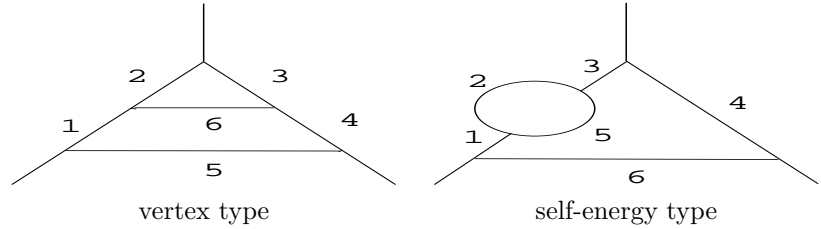

   \begin{center}
   \begin{tabular}{cc}
       {\def\epsfsize#1#2{0.5#1\epsfysize0.27#2}\epsfbox{gr3a.eps}} &
       {\def\epsfsize#1#2{0.5#1\epsfysize0.27#2}\epsfbox{gr5a.eps}} \\
        vertex type  & self-energy type\\
   \end{tabular} \end{center}
\caption{Assignment of Feynman Parameters}
\end{figure}

\section{Counter terms}
There are 58 diagrms including one loop counter term.
The contributions from these diagrams are  easily obtained,
by interpreting the counter term as the new coupling
including $(1/\ve)$. No divergence occurs in one
loop integration, however, we must keep order $\ve$
quantity in the integration to get final result.

\section{System of Calculation}
We first prepare the topology dependent part.
Corresponding to each topology, we can
assign the flow of loop momenta and external momenta.
Then we can calculate the important quatntity detU, $B_{ij}$
and the coefficients expressing the weight of
the flow of momenta $q$ and $p$ in each internal line by
using the conservation law of momentum at each vertex. These
are the prepared files.
The GRACE generates
diagrams and correponding FORM source codes.
By invoking FORM, the prepared files are included according to
their topology and FORTRAN source codes
giving  the UV and finite part contibution to $(g-2)$ are
generated. \\
Numerical integration over Feynman parameter space
 is done by BASES$^7$.
For test run, we have checked several diagrams.
We have reproduced the pure QED results by this system.
The UV divergence is extracted systematically.
The infrared$\;$(IR) singularity
is regulated by introducing a tiny photon mass.
The diagrams including self-energy type sub-diagram develop
superficial IR singularity due to the fact that the $x$-derivative
to get rid of UV part increases the power of denominator
function. We need subtle treatment for IR
singularity.

\textheight=7.8truein
\setcounter{footnote}{0}
\renewcommand{\thefootnote}{\alph{footnote}}

\nonumsection{Acknowledgements}
\noindent
The authors would like to thank Prof.T.Kinoshita for
introducing his formalism to them.
This work is supported in part by the Ministry of Education,
Science and Culture, Japan under the Grant-in-Aid for
International Scientific Research Program No.04044158.

\nonumsection{References}

\end{document}